\documentclass[preprints,review,accept,pdftex,moreauthors]{Definitions/mdpi} 

\usepackage{MnSymbol}
\usepackage{Definitions/abbreviations}

\firstpage{1} 
\pubvolume{1}
\issuenum{1}
\articlenumber{0}
\pubyear{2023}
\copyrightyear{2023}
\datereceived{ } 
\daterevised{ } 
\dateaccepted{ } 
\datepublished{ } 
\hreflink{https://doi.org/} 

\Title{Molecular gas kinematics in local early-type galaxies with ALMA}

\TitleCitation{Molecular gas kinematics in local early-type galaxies with ALMA}

\Author{Ilaria Ruffa$^{1,2\dagger}$\orcidA{} and Timothy A. Davis$^{1}$}

\AuthorNames{Ilaria Ruffa and Timothy A. Davis}

\AuthorCitation{Ruffa, I.; Davis, T. A.}

\address{%
$^{1}$ \quad Cardiff Hub for Astrophysics Research \&\ Technology, School of Physics \&\ Astronomy, Cardiff University, Queens Buildings, The Parade, Cardiff, CF24 3AA, UK\\
$^{2}$ \quad INAF - Istituto di Radioastronomia, via P.\ Gobetti 101, 40129 Bologna, Italy}

\corres{Correspondence: ruffai@cardiff.ac.uk; davist@cardiff.ac.uk}

\abstract{Local early-type galaxies (ETGs) are mostly populated by old stars, with little or no recent star formation activity. For this reason, they have historically been believed to be essentially devoid of cold gas, that is the fuel for the formation of new stars. Over the past two decades, however, increasingly-sensitive instrumentation observing the sky at (sub-)mm wavelengths has revealed the presence of significant amounts of cold molecular gas in the hearts of nearby ETGs. The unprecedented capabilities offered by the Atacama Large Millimeter/submillimeter Array (ALMA), in particular, have allowed us to obtain snapshots of the central regions of these ETGs with unprecedented detail, mapping this gas with higher sensitivity and resolution than ever possible before. Studies of the kinematics of the observed cold gas reservoirs are crucial for galaxy formation and evolution theories, providing - e.g. - constraints on the fundamental properties and fuelling/feedback processes of super-massive black holes (SMBHs) at the centre of these galaxies. In this brief review, we summarise what the first ten years of ALMA observations have taught us about the distribution and kinematics of the cold molecular gas component in nearby ellipticals and lenticulars.}


\begin{document}


\section{Introduction}\label{sec:intro}
High-quality imaging and spectroscopic surveys carried out over the past two decades (such as the Sloan Digital Sky Survey, SDSS; \citealp{York00}) have revealed the existence of a clear bi-modality in the colour and mass distribution of galaxies in the local Universe \citep[i.e.\,within $z<0.1$; e.g.][]{Baldry04}.\,One mode consists on late Hubble type (disc) galaxies characterised by significant ongoing star formation (and thus blue optical colours), stellar masses $M_{\ast}\lesssim10^{10.5}$~M$_{\odot}$ and a tight correlation between their star formation (SF) rate and $M_{\ast}$ \citep[known as ``star-forming main sequence''; e.g.][]{Schimi07}.\,The other population consists on the so-called early-type galaxies (ETGs), made of both ellipticals and lenticulars.\,These are generally dominated by old stars, which likely formed in bursts the time of their high-redshift progenitors, the so-called dusty star-forming galaxies (DSFG; also known as sub-millimetre galaxies, SMGs; see e.g.\,\citealp{Lapi18,Pantoni19,Giulietti23}).
There is very little or no ongoing SF in modern-day ETGs, and they have typical masses $M_{\ast}\gtrsim10^{10.5}$~M$_{\odot}$.\,Therefore, in classic optical colour-magnitude diagrams, ETGs give rise to what has been dubbed ``red - or  passive - sequence'' \citep[e.g.][]{Salim07}.

It is now widely believed that this bi-modality is a result of galaxy evolution \citep[e.g.][]{Faber07}.\,In this scenario, the transition of galaxies from blue, star-forming discs into giant ``red and dead'' spheroids is driven by the availability within the inter-stellar medium (ISM) of cold gas (i.e.\,the fuel for star formation), which is in turn regulated by the balance between gas accretion and depletion processes \citep[e.g.][]{Lilly13}.This transition occurs relatively quickly (as only a very minor fraction of galaxies populates the region between the two sequences, the so-called “green valley") and - once completed - it is permanent (at least in the majority of cases). Galaxy formation and evolution theories still struggle to reproduce the observed properties of red sequence galaxies: how they are kept as passive spheroids while growing in mass up to $M_{\ast}\sim10^{12}$~M$_{\odot}$, how the fuelling/feedback processes of their central super-massive black holes (SMBHs) are powered and sustained, are only some subjects of long-standing debates \citep[e.g.][]{Kormendy13,Hardcastle18,Donofrio21}.

Due to the optical properties described above, local ETGs have historically been believed to be essentially devoid of cold gas. Over the past twenty years, however, increasingly-sensitive millimetre-wave instruments (such as the Institut de radioastronomie millimétrique 30m single-dish, IRAM-30m; the Northern Extended Millimetre Array, NOEMA; and the Combined Array for Research in Millimeter-wave Astronomy, CARMA) revealed that molecular gas is present in at least $\approx$25\% of local ETGs \cite{Knapp1996, Welch2003, Combes2007, Sage2007, Young2008, Welch2010, Young2011, Crocker2011, Alatalo2013,Davis2019}, and this detection rate is insensitive to galaxy mass (but potentially not to the dynamical state of the system; \citealp{Young2011,Davis2019}). Over the past decade, in particular, the resolution and sensitivity provided by the Atacama Large Millimeter/submillimeter Array (ALMA) allowed us to obtain snapshots of their cores with unprecedented detail and to accurately map the distribution and kinematics of this gas, thus opening entirely new avenues in the study of the cold ISM in this type of objects \citep[see e.g.][and references therein]{Morganti15a,Tremblay16,Davis17,Boizelle17,Maccagni18,Harrison18,Smith19,North19,Sansom19,Olivares19,Ruffa19a,Ruffa19b,Rose19,Matsui19,North21,Smith21a,Boizelle21,Young21,Ruffa22,Young22,Glass22,Temi22,Torresi22,Ruffa23,Rose23,Audibert23,Maccagni23,Williams23,Elford24,Dominiak24a,Dominiak24b,Rose24}.

The observation of significant amounts of cold gas in red sequence galaxies triggered the idea that also these objects experience cyclical regenerations of their gaseous reservoirs, which must be then retained for a while within their deep potential wells. How cold gas re-forms and settles down in local ETGs, however, is still an open question. According to those identified as the most plausible scenarios, it may be either internally re-generated (predominantly through hot halo cooling) or externally accreted (through interactions or minor mergers). Evidence is mounting that the large-scale environment where the galaxy is located plays a crucial role in determining the preferred among these mechanisms \citep[see e.g.][for extensive discussions in this regard]{Davis19,Storchi19,Ruffa19a,Ruffa19b,Ruffa22,Temi22,Maccagni23}.

Massive spheroids are also the preferential hosts of jetted active galactic nuclei (AGN), which are those producing (relatively) strong kinetic feedback in the form of collimated, relativistic outflows of non-thermal plasma (i.e.\,the radio jets; \citealp[see e.g.][]{Padovani16}).\,This AGN category currently includes all types of radio galaxies (RGs), the most powerful low-ionization nuclear emission-line regions (LINERs) and some quasars/Seyferts \citep[e.g.][]{Heckman14,Padovani17a}. The first decade of ALMA observations enabled detailed studies of the cold gas reservoirs in each of these jetted-AGN types, helping to improve our understanding of their powering mechanism(s), as well as of the complex interplay between radio jets and the surrounding gaseous medium at all spatial scales \citep[from the inner sub-kpc to tens of kpc; e.g.][]{Garcia14,Dasyra16,Oosterloo17,Ruffa19a,Ruffa20,Ruffa22,Audibert23,Dotti24}. In this context, a particularly interesting example is the one of low-excitation radio galaxies (LERGs), which are the most common type of RGs at $z<0.1$. LERGs are typically hosted by the most massive ETGs (with $M_{\ast}>10^{11}$~M$_{\odot}$; \citealp{Best12}), whereby the central SMBH accretes gas at very low rates ($<<1$\% of the Eddington limit) and produces almost exclusively kinetic feedback \citep[e.g.][]{Best12,Heckman14,Ruffa19a,Ruffa19b}. Due to their host-galaxy properties, the SMBHs in LERGs have traditionally been believed to be fuelled by direct inefficient accretion of hot X-ray emitting gas from the  circum-galactic and/or inter-galactic medium \citep[CGM and IGM, respectively; see e.g.][]{Hardcastle07}.\,The detection of large masses of cold gas at their centres, however, started to pose this picture into question, providing new fundamental constraints on the powering mechanism of LERGs \citep[see e.g.][]{Ruffa19a,Ruffa19b,Ruffa22,Ruffa23}. 


In this manuscript, we present a short review on what we have learnt in the past ten years by studying the distribution and kinematics of the molecular gas in local ETGs when observing them at very high resolution and sensitivity with ALMA.
\section{Molecular gas in early-type galaxies}\label{sec:mol_gas_in_ETG}
Carbon Monoxide (CO) is the most used molecular gas tracer: it is the most abundant molecule after H$_{2}$ (which is rather difficult to detect in emission; see e.g.\,\citealp{Bolatto13,Zovaro19}), and has low requirements in terms of both excitation temperature (i.e.\,$T_{\rm ex}=5.53$~K for the ground $J =1\rightarrow0$ rotational transition) and critical density ($n_{\rm crit}$; i.e.\,the density at which collisional excitation balances spontaneous radiative de-excitation, which for CO is $\gtrsim2200$~cm$^{-3}$). All this makes CO relatively easy to detect in extragalactic sources \citep[e.g.][]{Carilli13}. Other molecules, such as Formylium (HCO$^{+}$), Carbon Monosulfide (CS), and Hydrogen Cyanide (HCN), are instead the most used tracers of the dense ($n_{\rm crit}>10^{4}$~cm$^{-3}$) molecular gas component \citep[e.g.][]{Topal16,Bigiel16,Young22}. The rotational transitions of all these lines up to $J_{\rm upper}=8-10$ (up to $J_{\rm upper}=19$  for CS) have rest frequencies between $\approx100$~GHz and $\approx900$~GHz, making them ideal for ALMA observations in nearby galaxies. An in-exhaustive search on the ALMA Science Archive carried out at the time of writing finds available observations of different CO lines (mostly of the transitions with $J_{\rm upper} = 1$, $2$ or $3$) and/or dense molecular gas tracers for about 150 single ETGs within $z\lesssim0.05$. Of these, about 90 have been published so far, and about 70 clearly detected (at least) in CO. In the following, we summarise the main properties in terms of gas distribution and kinematics that have been inferred from these studies. We note that, for clarity, in this review we divide the ETG population into three different classes: normal ETGs (standard, passive spheroids without any clear signs of nuclear activity), jetted-AGN hosts, and brightest central galaxies (BCGs; i.e.\,very massive ETGs at the centre of rich groups and clusters, typically hosting also a jetted-AGN).

\subsection{Cold gas content and distribution}\label{sec:gas_distribution}
Advances in millimetre instrumentation mean that we are now able to detect and map the gas reservoirs in ETGs, even in systems with low gas masses.\,The typical molecular gas masses found in  ETGs studied to date range from $\sim10^{7}$ to $\sim10^{8}$~M$\odot$, but can even go up to $\sim10^{9}-10^{10}$~M$\odot$ in some exceptional cases \citep[see e.g.][]{Morganti15a,Voort18,Ruffa19a,Smith21a}. We caution, however, that these masses can have relatively large uncertainties, mainly introduced by the adopted CO-to-H$_{2}$ conversion factor \citep[see][for a comprehensive review on this subject]{Bolatto13} and the ratio over the ground CO transition (when using higher-J CO lines for the estimation; see e.g.\,\citealp{Ruffa19a,Ruffa22}, for discussions in this regard). 

\begin{figure}[]
\begin{adjustwidth}{-\extralength}{0cm}
\centering
\includegraphics[clip=true, trim={1 1 1 1}, scale=0.85]{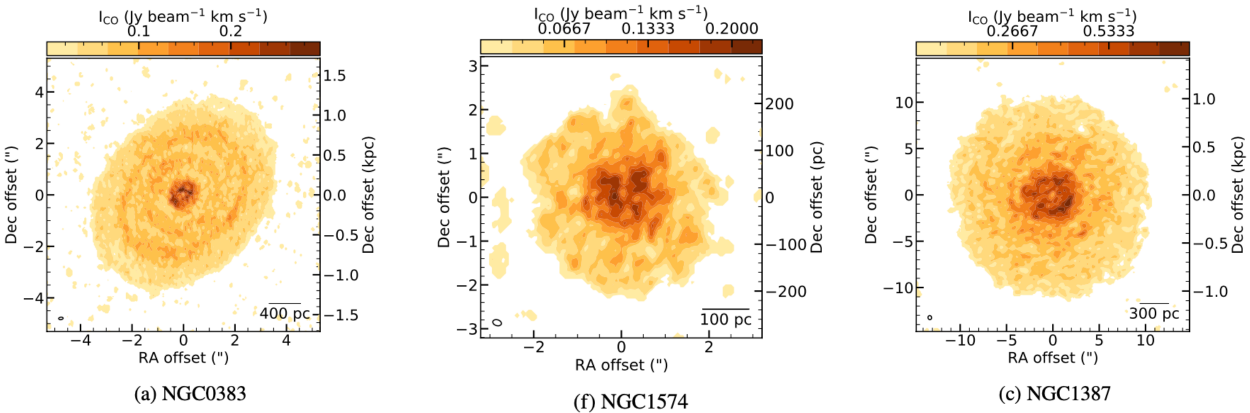}
\end{adjustwidth}
\caption{Integrated-intensity (i.e.\,moment 0) maps of the CO(2-1) transition observed at high-resolution with ALMA in three local ETGs (as labelled at the bottom of each panel). These are taken as an example of normal ETGs and (non-BCG) jetted-AGN hosts which are characterised by compact and smooth disc-like molecular gas structures extending around the central galaxy regions (see the text for details). The synthesised beam and scale bar are shown in the bottom-left and bottom-right corners, respectively, of each panel. The physical resolution reached is similar in each case ($\approx30$~pc). The bar on top of each panel shows the colour scale in Jy~beam$^{-1}$~km~s$^{-1}$, and coordinates are given as relative to the image phase centre. The maps are reproduced from \citealt{Davis22}.\label{fig:CO_morphology}}
\end{figure}

In normal ETGs and jetted-AGN hosts (including a few BCGs), the detected CO emission is usually observed to be distributed in thin disc-like structures, extending on scales from a few hundred parsecs to a few kpc around the nuclear regions of the host galaxy (see Figure~\ref{fig:CO_morphology}). Such compact molecular gas morphologies are also typically very smooth (even when observed at spatial resolutions as high as 10 pc, as opposed to what found in typical late-type galaxies; \citealp[see e.g.][]{Davis22}), and have been interpreted as an indication of gas that has mostly settled into the potential well of the host galaxy. 

\begin{figure}[]
\begin{adjustwidth}{-\extralength}{0cm}
\centering
\includegraphics[clip=true, trim={1 1 1 1}, scale=0.68]{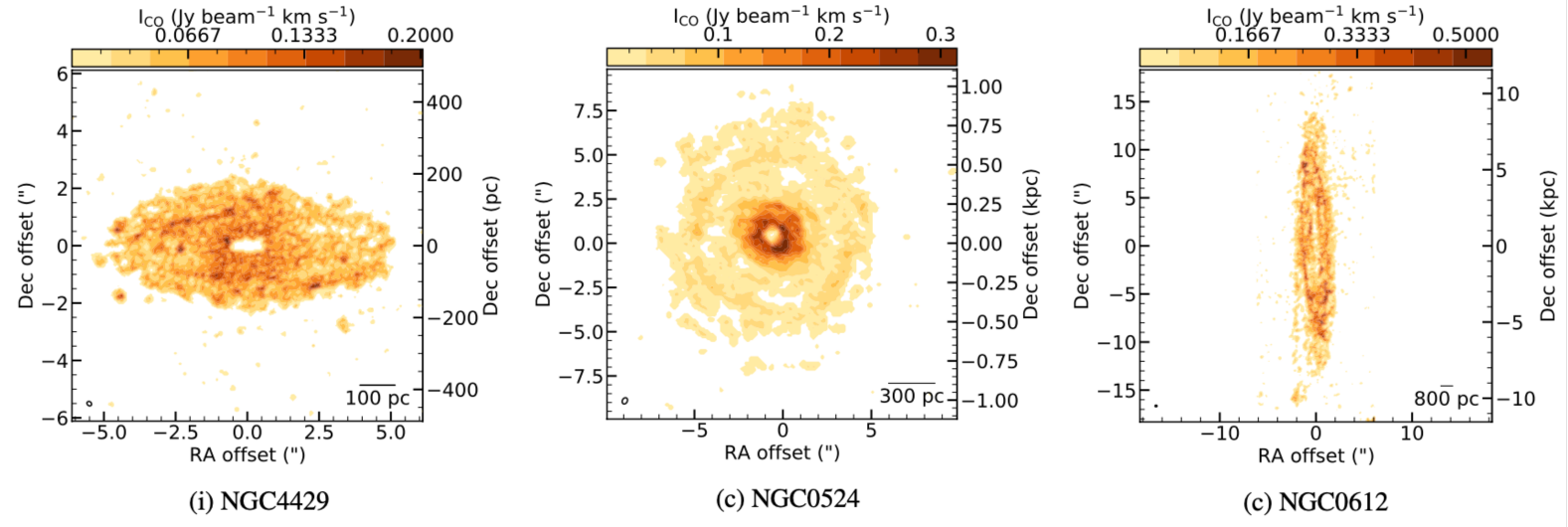}
\end{adjustwidth}
\caption{Integrated-intensity (i.e.\,moment 0) maps of the CO(2-1) transition observed at high-resolution with ALMA in three local ETGs (as labelled at the bottom of each panel). These are taken as an example of normal ETGs and (non-BCG) jetted-AGN hosts which are characterised by evident nuclear gas deficiencies or proper ring-like shapes. The synthesised beam and scale bar are shown in the bottom-left and bottom-right corners, respectively, of each panel. The physical resolution reached is similar in each case ($\approx30$~pc). The bar on top of each panel shows the colour scale in Jy~beam$^{-1}$~km~s$^{-1}$, and coordinates are given as relative to the image phase centre. The maps are reproduced from \citealt{Davis22}.\label{fig:CO_morphology_holes}}
\end{figure}

Nuclear gas deficiencies or proper central holes (i.e.\,ring-like shapes) are also observed in about 20\% of the cases, including both normal ETGs and jetted-AGN hosts \citep[e.g.][]{Alatalo2013,Davis13,Combes19,Smith19,Garcia19,Ruffa19a,Smith21a} (see Figure~\ref{fig:CO_morphology_holes}). The sizes of such circumnuclear cavities usually ranges from few tens to few hundreds of parsec and their origin is still puzzling. Some mechanism may either dissociate the molecular gas or prevents it from forming (or accumulating) in the very centre of the galaxy. This can occur in presence of an intense radiation field, such as the one due to strong X-ray radiation from a radiatively-efficient (quasar-like) AGN (producing the so-called X-ray dissociation regions, XDRs). Various studies suggest that XDRs are particularly effective in dissociating molecules such as HCO$^{+}$ \citep[see e.g.][]{Imanishi14,Imanishi18}. On the other hand, in radiatively-inefficient (LINER-like) AGN, the X-ray radiation arising from the accretion process is much less extreme in terms of dissociating power. Therefore, in cases like these, the molecular gas at the centre can simply be much denser and excited towards high-$J$ transitions due to X-ray heating or shocks and compression induced by an interaction with the expanding radio jets (see Section~\ref{sec:gas_kinematics} for details). When this occurs, the central gap is usually filled when observing CO lines with $J_{\rm upper}\geq3$ \citep[e.g.][]{Wada18,Combes19,Ruffa22} and/or dense molecular gas tracers \citep[e.g.][]{Imanishi18}. Prominent molecular gas outflows (due to either jet-induced or radiative AGN feedback) have been also indicated as particularly plausible processes for the removal of gas from the circumnuclear regions \citep[e.g.][see also \citealp{Harrison24} for a recent review on the effects of AGN feedback on the circumnuclear gas reservoirs]{Garcia14, GarciaBurillo21}. 

Dynamical effects can also affect the distribution and survival of molecular gas in those areas, thus favouring the formation of nuclear cavities. For instance, the strong shear or tidal acceleration expected near the SMBH can disrupt gas clouds. 
Non-axisymmetric gravitational instabilities induced by stellar bars can also give rise to nuclear gaps, relying on bar-induced gravitational torques that force the gas outwards to the inner Lindblad resonance (ILR), or inwards on to the central SMBH \citep[e.g.][]{Combes01}. Stellar bars, however, give rise also to characteristic features in molecular gas velocity curves (i.e.\,position-velocity diagrams, PVDs), especially in galaxies with high inclination angles with respect to the line-of-sight (i.e.\,$\theta_{\rm inc}\gtrsim60^{\circ}$). In these cases, molecular gas PVDs typically show X-shapes or sharp edges near the turnover radius (i.e.\,the radius that marks the transition between the central rising and the external flat part of the velocity curve; see Section~\ref{sec:gas_kinematics}). These are both clear signatures indicative of the presence of two velocity components along the line-of-sight: an inner rapidly rising component associated with gas located within the ILR, and an outer slowly rising component associated with gas on nearly circular orbits beyond co-rotation \citep[e.g.][]{Alatalo13,Combes13,Voort18}. Based on all the above, it is clear that high-quality, multi-wavelength observations of the various galaxy components (e.g.\,stars, radio jet, etc), in conjunction with those of multiple molecular gas transitions, are key to discriminate between different mechanisms and draw solid conclusions on the origin of the observed circumnuclear molecular gas holes. 

\begin{figure}[]
\begin{adjustwidth}{-\extralength}{0cm}
\centering
\includegraphics[clip=true, trim={1 1 1 1}, scale=0.68]{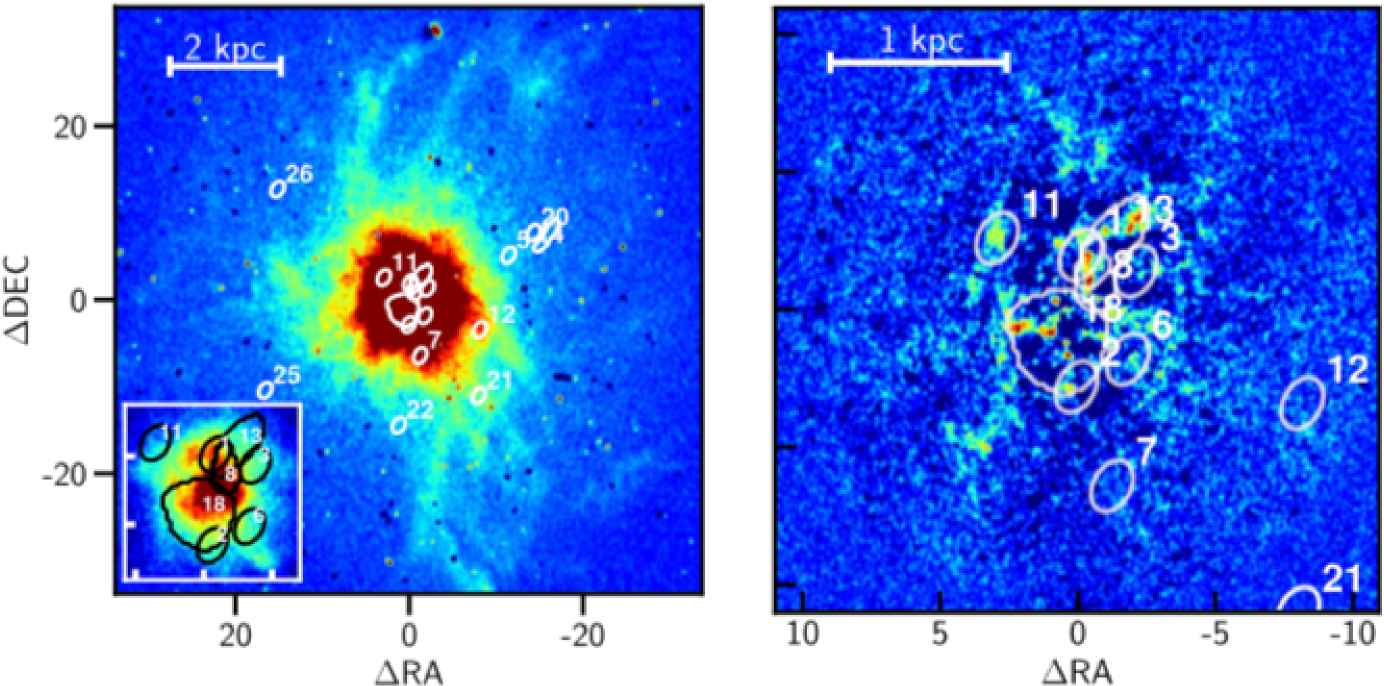}
\end{adjustwidth}
\caption{Optical H$\alpha+$[NII] integrated intensity map (left panel) and dust absorption map (right panel) of the BCG NGC\,5044 with overlaid in black (in the zoom-in inset on the left panel) and white the contours of the 17 CO(2-1) clouds detected with ALMA. The colour sequence (going from blue to red) indicates increasing values of ionised gas intensities and dust absorption. The maps are reproduced with permission from \citealt{Temi18}. See the text for details. \label{fig:NGC5044}}
\end{figure}

The distribution of molecular gas in BCGs is often 
different from those described above, with large amounts of cold gas typically detected in the form of kpc-scale filamentary or blob-like structures \citep[e.g.][]{David14,Werner14,Russell16,Temi18,Tremblay18,Nagai19,Matsui19,Olivares19,North21,Temi22}. These are thought to be cooling from the hot, X-ray-emitting intra-cluster (or intra-group) medium, as described - for instance - in the framework of chaotic cold accretion models (CCA; see \citealp[e.g.][]{Gaspari13,King15,Gaspari15,Gaspari17}). According to CCA, thermal instabilities in the hot halo lead to the formation of decoupled dense gas structures, being distributed into clouds and filaments while cooling down to $T\ll10^{3}$~K. Stochastic dissipative collisions between such cold gas structures then significantly reduce their angular momentum, causing them to chaotically “rain” towards the central SMBH. Recent works demonstrate that relatively tight correlations exist between the molecular gas mass and the hot halo properties (i.e.\,temperature, mass and luminosity) of group-/cluster-centered ETGs \citep{Babyk18}, providing further evidence of a close hot-cold gas connection in these objects. A notable case among this category is the one of NGC~5044, a well-known ETG at the centre of an X-ray bright group observed multiple times with ALMA in CO(2-1) \citep[][]{David14,Temi18} (Figure~\ref{fig:NGC5044}). These data reveal the presence of 17 cold gas clouds within the central $\sim6$~kpc of the galaxy, each extending on scales of less than  $300$~pc. Interestingly, these clouds appear to be mostly co-spatial with the warm gas phase (as visible in the H$\alpha+$[N {\sc II}] map shown in the left panel of Figure~\ref{fig:NGC5044}) and with the X-ray emission \citep{Werner14}, resulting in the multi-phase cooling gas distribution predicted in CCA simulations \citep{Gaspari17}.

\subsection{Molecular gas kinematics}\label{sec:gas_kinematics}
Quantitative details on the dynamical state of the molecular gas in ETGs are typically obtained from the 3D kinematic modelling of spatially-resolved gas distributions (using publicly-available tools such as the Kinematic Molecular Simulation tool, {\sc KinMS}; \citealp[]{Davis13}, and $^{\rm 3D}$BAROLO; \citealp[]{DiTeodoro15}) and/or from the harmonic decomposition of 2D line-of-sight gas velocity maps (using packages such as {\sc kinemetry}; \citealp{Kraj06}). Qualitatively, information on the gas kinematics can be also inferred by-eye from the integrated line profiles and the classic 3D imaging data products, such as the mean line-of-sight velocity map (moment 1) and the PVDs. In the following, we briefly summarise the results obtained from these type of analyses, splitting the discussion of normal ETGs and jetted-AGN hosts (including BCGs). We note that maps of the mean line-of-sight velocity dispersion (moment 2) are also typically used to get spatially-resolved information on the line velocity broadening, and thus on the corresponding gas dynamical state. We caution, however, that moment maps of order higher than 1 are the most severely affected by the signal-to-noise ratio (S/N) of the line detection (due to the way they are defined; see e.g.\,\url{https://spectral-cube.readthedocs.io/en/latest/moments.html#linewidth-maps}), becoming essentially meaningless when this is too low. Furthermore, line-of-sight velocity dispersions can be significantly inflated due to observational effects. Among these, beam smearing (i.e.\,contamination from partially resolved or unresolved velocity gradients within the host galaxy) usually dominates in regions of large velocity gradients, such as the inner regions of galaxies.\,For this reason, it is highly recommended to always validate the observed gas velocity dispersion ($\sigma_{\rm gas}$) by means of a full 3D kinematic modelling which takes into account these sorts of observational effects (and thus provide an estimate of the intrinsic $\sigma_{\rm gas}$), before drawing strong conclusions \citep[see e.g.][]{Davis17,Ruffa23}. The values of $\sigma_{\rm gas}$ discussed in the following can be all considered as intrinsic. 

\subsubsection{Normal ETGs}
\begin{figure}[]
\begin{adjustwidth}{-\extralength}{0cm}
\centering
\includegraphics[clip=true, trim={1 5 1 5}, scale=0.63]{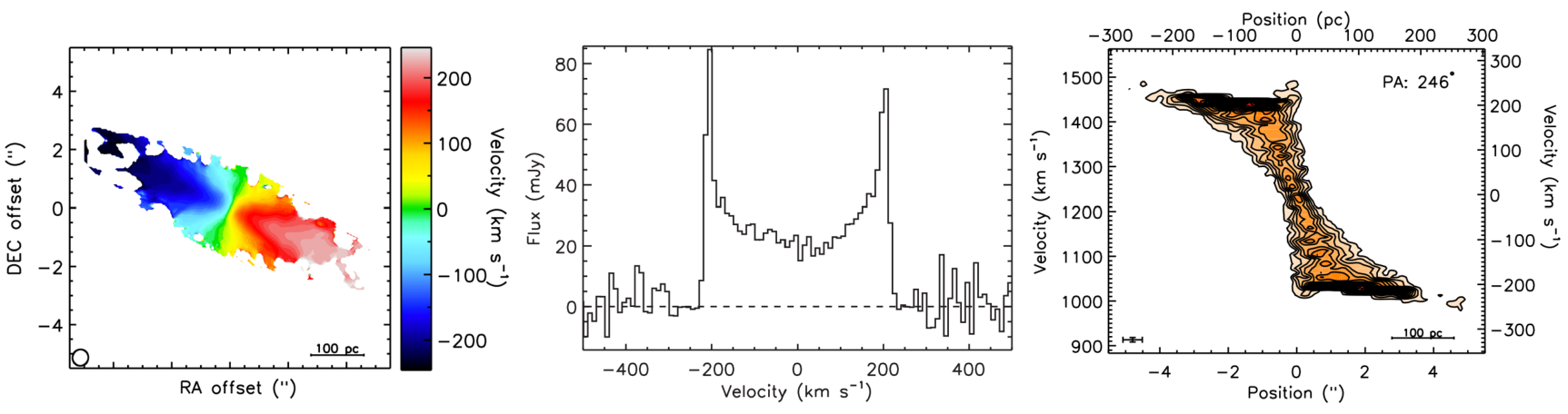}
\end{adjustwidth}
\caption{CO(2-1) mean line-of-sight velocity (moment 1) map (left panel), integrated spectral profile (middle panel), and major axis position-velocity diagram (PVD; right panel) of the normal ETG NGC\,4697. All the maps are reproduced with permission from \citet{Davis17}. In the left panel, the synthesised beam is shown in the bottom-left corner, and the colour bar to the right show the colour scales in km~s$^{-1}$. A scale bars is also shown in the bottom-right corner of the left and right panels. Coordinates are given as offsets (in arcseconds) relative to the image phase centre. \label{fig:CO_kin}}
\end{figure}

Studies of molecular gas kinematics (mostly based on ALMA observations of the CO(2-1) transition) show that in normal ETGs the gas is regularly rotating. Any departure from symmetry (such as inclination or velocity warps) and secondary kinematic components (such as inflow/outflow motions) are negligible or completely absent in the majority of these cases \citep[e.g.][]{Topal16,Davis17,Boizelle17,Smith19,Boizelle19,Kabasares22,Kabasares24,Dominiak24a}. For instance, by analysing the dynamical state of the molecular gas in five local normal ETGs observed at high-resolution ($\leq$100~pc) in CO(2-1) with ALMA, \citet{Boizelle17} found that in all the targets deviations from models assuming gas in pure circular motions were very small ($\lesssim10$~km~s$^{-1}$) and warping of the gas discs of low magnitude ($\lesssim10^{\circ}$). This is also consistent with earlier results from the ATLAS\textsuperscript{3D} survey \citep{Cappellari11}, which were obtained by studying the molecular gas kinematics of $\approx30$ normal (gas-rich) ETGs observed in CO(1-0) with CARMA \citep[see][]{Alatalo13,Davis13}. These findings have been generally interpreted as an indication of gas that is both dynamically and morphologically relaxed, and thus fully settled into the potential well of the host galaxy. In cases like these, the mean line-of-sight velocity maps and integrated line profiles respectively show fairly regular/smooth velocity gradients and prominent, nearly-symmetric double-horned shapes characteristic of rotating bodies [see Figure~\ref{fig:CO_kin}, left and middle panel, respectively]. The PVDs extracted along the major axis of such regularly-rotating gas distributions are usually characterised by a steep velocity gradient at the centre, followed by a velocity flattening on larger scales [see Figure~\ref{fig:CO_kin}, right panel]. These PVDs also do not present any clear asymmetry and/or secondary kinematic component (with the exception, in well-resolved cases such as the one shown in Figure~\ref{fig:CO_kin}, of the Keplerian upturn at the very centre of the gas distribution; see Section~\ref{sec:dyn_BH_masses} for details). In these cases, the typical mean gas velocity dispersion is in the range $\sigma_{\rm gas} \sim 10-20$~km~s$^{-1}$ (where these are values found by averaging throughout the disc/ring; \citealp[see e.g.][]{Davis18,Voort18,Ruffa19b,Ruffa23}). As a consequence, the molecular gas in normal ETGs is consistent with being dynamically cold and the disc/ring fully rotationally supported (with typical $v_{\rm rot}/\sigma_{\rm gas}$ ratios $\gtrsim10$, where $v_{\rm rot}$ is the de-projected circular velocity of the gas at the turnover radius; \citealp[see e.g.][]{Wisnioski15}).

\subsubsection{Jetted-AGN hosts and jet-ISM interactions}
\begin{figure}[]
\begin{adjustwidth}{-\extralength}{0cm}
\centering
\includegraphics[clip=true, trim={1 5 1 2}, scale=0.63]{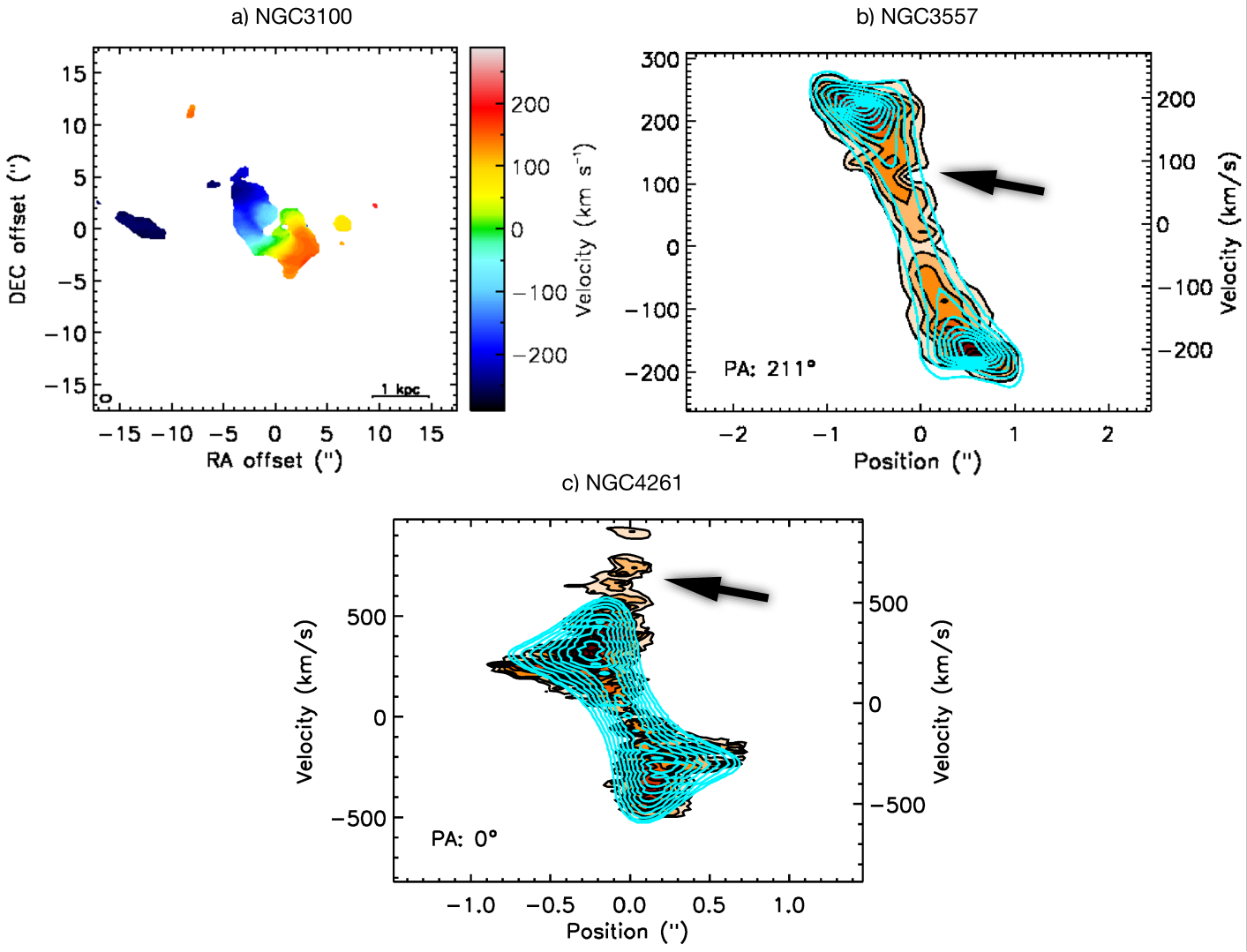}
\end{adjustwidth}
\caption{CO(2-1) mean line-of-sight velocity (moment 1) map of NGC3100 (top-left panel) and major axis PVDs of NGC3557 and NGC4261 (top-right and bottom panels, respectively), all taken as examples of jetted-AGN hosts and reproduced from \citet{Ruffa19b,Ruffa23}. In the top-right and bottom panels, the contours of the best-fitting kinematic model assuming gas in pure circular motions are overlaid in cyan, clearly highlighting the presence of unrelaxed gas sub-structures which are not reproduced by such simple axisymmetric model (see \citealt{Ruffa19b,Ruffa23} for details). The black arrows points at the location of such asymmetries. In the top-left panel, the synthesised beam and scale bar are shown in the bottom-left  and bottom-right corners, respectively. The colour bar to the right show the colour scales in km~s$^{-1}$. Coordinates are given as offsets (in arcseconds) relative to the image phase centre. \label{fig:jetted_AGN_kin}}
\end{figure}

Fully relaxed CO discs/rings are not a standard feature of all local ETGs. Indeed, a number of works have now demonstrated that clear differences can be observed in the molecular gas kinematics of jetted-AGN hosts \citep[e.g.][]{Garcia14,Dasyra16,Oosterloo17,Ruffa19b,Pap23,Audibert23,Ruffa23}. For instance, \citet{Ruffa19b,Ruffa22,Ruffa23} carried out the 3D kinematic modelling of the CO(2-1) transition in seven nearby ETGs hosting LERGs. In all the cases, they found that - while the bulk of the gas can be still considered as regularly rotating and dynamically cold - localised distortions in both the morphology and kinematics are ubiquitously observed. These cannot be reproduced by simple axisymmetric models assuming gas in pure circular motion, thus indicating the presence of unrelaxed gas sub-structures (a small compilation of these examples is presented in Figure~\ref{fig:jetted_AGN_kin}). In particular, multiple kinematic components and/or warps (i.e.\,tilted or s-shaped velocity iso-contours) can be identified in almost all the cases, the former with velocities up to $\approx600$~km~s$^{-1}$ (see e.g.\,the lower panel of Figure~\ref{fig:jetted_AGN_kin}) and the latter with magnitudes $\gtrsim40^{\circ}$ (see e.g.\,the upper-left panel of Figure~\ref{fig:jetted_AGN_kin}). Similar features can arise from a number of different physical processes, such as stellar bars, settling mechanisms and inner spiral perturbations, without necessarily invoking AGN feedback \citep[see e.g.][for comprehensive discussions in this regard]{Ruffa19b,Ruffa23}. However, the fact that such disturbances are systematically observed in ETGs hosting AGN with active radio jets (as opposite to what found in their inactive counterpart; see above) suggests that there might be a correlation between the two \citep[see e.g.][]{Ruffa20}. A detailed discussion on the role of radio jets in modifying the physics and kinematics of the molecular gas on the (sub-)kpc scales of nearby active galaxies is beyond the scope of this work. However, in the following we briefly review the notions and findings that are most relevant in this framework.

\begin{figure}[]
\centering
\includegraphics[clip=true, trim={1 5 1 5}, scale=0.62]{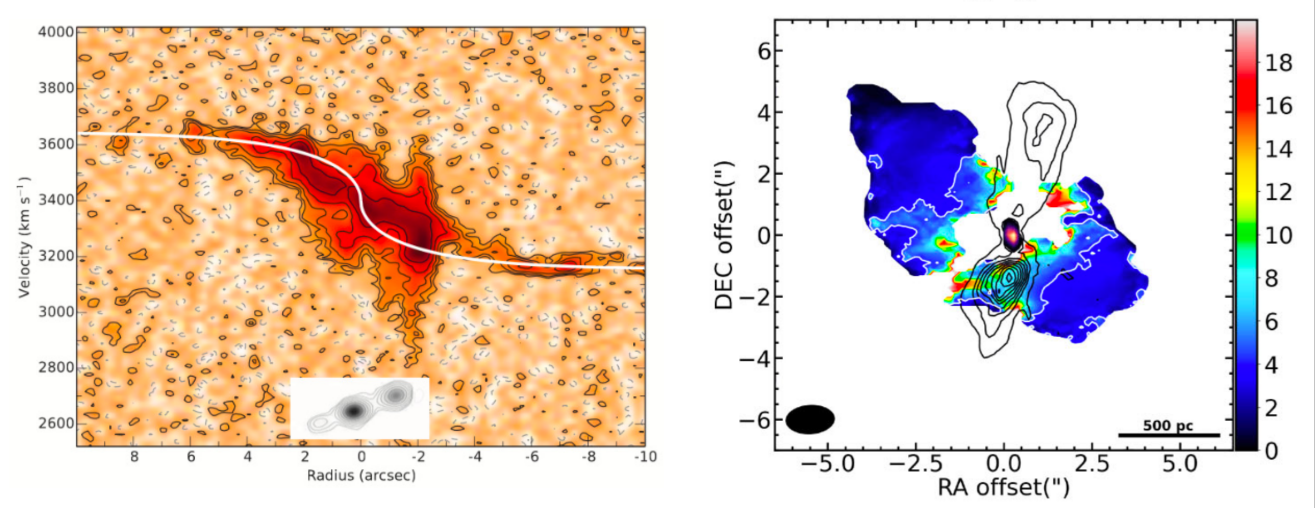}
\caption{{\bf Left panel:} Major-axis PVD of the CO(2-1) line emission observed with ALMA in the jetted-AGN host IC\,5063. The image is reproduced with permission from \citet{Morganti15b}. The white line represents the rotation curve derived from {\it Hubble Space Telescope} (HST) photometry and illustrates the expected kinematics of gas in regular circular rotation. The inset in the central-bottom region shows the position and morphology of the core-double lobe radio structure as seen in the ALMA continuum at 230~GHz. {\bf Right panel:} $R_{\rm 21} \equiv S_{\rm CO(2-1)}/S_{\rm CO(1-0)}$ map of NGC\,3100 with 10~GHz continuum contours of the core-double lobe radio structure overlaid in black. The map is reproduced from \citet{Ruffa22}. The white contour roughly marks the transition between the outer low-excitation component with $R_{\rm 21}<4$ and the inner high-excitation component. The bar to the right shows the colour scale. The beam size and scale bar are shown in black in the bottom-left and bottom-right corners, respectively. Coordinates are given as relative to the image phase centre; East is to the left and North to the top.\label{fig:CO_jet}}
\end{figure}

The idea that expanding radio jets can have an impact on the gaseous reservoirs on (sub-)kpc scales is relatively recent. Indeed, up until some time ago, they were believed to operate mostly on tens of kpc scales (by inflating cavities in the hot X-ray emitting atmospheres of galaxies, groups and clusters; \citealp[see e.g.][]{McNamara12}). Substantial recent work on 3D hydrodynamical simulations of jet-ISM interactions demonstrated that these can actually often occur on the circumnuclear regions of jetted-AGN hosts, altering the gas physics, distribution and kinematics at various levels \citep[see][and references therein]{Wagner12,Wagner16,Mukherjee18a,Mukherjee18b,Cielo18,Mukherjee21}. These works indeed show that, while driving their way outward, the relativistic jet plasma can produce energy-driven bubbles that expand through the surrounding medium. The expanding bubbles can ablate the molecular gas clouds (i.e.\,locally compressing and/or fragmenting the gas distribution), accelerating them up to $\approx$1000~km~s$^{-1}$ over a wide range of directions with respect to the jet axis and inducing shocks and turbulence into the cold molecular medium (thus yielding to higher gas excitation conditions; see also Section~\ref{sec:gas_distribution}). Over the past decade, thanks in particular to the advent of ALMA, such theoretical predictions found support in a few confirmed examples of (sub-)kpc scale jet-iSM interactions \citep[see][]{Garcia14,Dasyra15,Oosterloo17,Murthy19,Zovaro19,Ruffa20,Ruffa22,Pap23,Audibert23}. A remarkable case in the context of this review is the one of the nearby ($z=0.011341$) lenticular galaxy IC\,5063, hosting the young (i.e.\,$t_{\rm age}\lesssim1$~Myr) core-double lobe radio source PKS 2048-57 and observed at high resolution with ALMA in different molecular gas transitions \citep[see][]{Morganti15a,Dasyra16,Oosterloo17}. These observations demonstrate a prominent alteration of the molecular gas kinematics in the regions co-spatial with the radio lobes, with anomalously high gas velocities (up to 650~km~s$^{-1}$) that can be unequivocally explained as a jet-driven outflow with a mass rate of $\sim12$~M$_{\odot}$ [see Figure~\ref{fig:CO_jet}, left panel]. Molecular line ratios are also powerful tools to investigate the physical conditions of the gas in various environments, as different molecules, isotopologues, and transitions of the same species trace different gas components within the same galaxy. The ratios of the various molecular transitions observed in IC\,5063 allow to demonstrate that the molecular gas physics is modified around the radio source, varying from optically-thick and sub-thermally excited in the unperturbed gas to optically-thin with a high excitation temperature ($>50$~K) in the regions affected by the jet-driven outflow. Signs of gas fragmentation are also found, suggesting that the outflowing molecular gas is clumpy. Another interesting (although less extreme) example is the one of the radio galaxy PKS 0958–314, hosted by the local ($z=0.0088$) lenticular galaxy NGC\,3100 and detected at high-resolution with ALMA in multiple CO transitions (up to $J_{\rm upper}=3$; see \citealp{Ruffa19b,Ruffa22}). In this object, the ongoing jet-molecular gas coupling is inducing an outflow in the plane of the CO distribution with $v_{\rm max}\approx 200$~km~s$^{-1}$ and $\dot{M}\lesssim 0.12$~M$_{\odot}$~yr$^{-1}$. The interaction is also strongly altering the molecular gas physics by inducing very high CO line ratios and thus high gas excitation (with temperatures $T_{\rm ex}\gtrsim50$~K) around the nuclear regions [see Figure~\ref{fig:CO_jet}, right panel]. Other examples of local ETGs where it is confirmed that the presence of active radio jets is somehow altering the physics and kinematics of the molecular gas reservoirs on (sub-)kpc scales are NGC\,1266 \cite{Alatalo11,Nyland2013}, NGC\,1167 \citep{Murthy22} and NGC\,6328 \citep{Pap21,Pap23}.

The extent of jet-induced perturbations is predicted to vary depending mainly on the radio jet power and relative jet–gas orientation. In this scenario, jets of intermediate power (P$_{\rm jet}\sim10^{43}-10^{44}$~erg~s$^{-1}$) at large ($\geq45^{\circ}$) angles to the gas disc are predicted to interact more strongly and for longer times than jets of higher power (P$_{\rm jet}>10^{44}$~erg~s$^{-1}$) oriented perpendicularly to the disc plane \citep{Mukherjee16,Mukherjee18a,Mukherjee18b}. It has been also suggested that the strongest jet-ISM coupling (and thus the biggest impact onto the gas physics {\it and} kinematics) occurs in the very early phases of the jet evolution (i.e.\,within less than 1 Myr from their launch), while the jets are still ``trapped'' within the dense nuclear gas layers \citep{Mukherjee18a,Morganti21}. The most extreme cases of the jet-ISM interactions mentioned above (and the few others in non early-type hosts; see e.g.\,\citealp{Zovaro19,Audibert23}) have indeed been all caught in such an evolutionary phase of the jet, supporting this scenario. Once the jets break out from the dense nuclear regions and expand on larger scales, their impact should become progressively less extreme and most of the gas affected by their transit is expected to eventually falls back into the central regions, settling into the galaxy potential on estimated timescale of the order of tens of Myr \citep[e.g.][]{Mukherjee16}. This would explain why in presence of a 2 Myr-old radio source like the one hosted in NGC\,3100 the jet-induced perturbations in the gas kinematics are much less extreme than in objects like IC\,5063. At the same time, this picture makes plausible that - when low-level disturbances compatible with a jet-ISM interaction are observed in systems with mature (large-scale) radio jets - we are actually observing the residual effects of a strong coupling occurred in the past, with the gas still in the process of settling back and relaxing into the potential well \citep[see e.g.][]{Ruffa20}. The many details of these processes, however, have yet to be explored and thus remain unclear. Large samples of objects covering wide ranges of jet ages, powers, and jet–ISM relative orientation are required to gain a better understanding of the incidence and properties of jet–ISM interactions. This is essential information to fully assess the role of jet-induced feedback in galaxy evolution \citep{Morganti21}.

All of the properties illustrated above apply also to the few BCGs where the molecular gas has been observed to be settled in discs/rings onto the central (sub-)kpc scales \citep[see e.g.][]{Ruffa19a,Rose24}. On the other hand, however, kpc-scale cold gas blobs and filaments associated with hot gas cooling flows and observed in the majority of BCGs are both predicted \citep[e.g.][]{Gaspari17} and observed \citep[e.g.][]{David14,Russell16,Tremblay18,Temi18,Olivares19,Oosterloo24} to show very complex velocity patterns, with little or no evidence of regular rotation. This has been generally interpreted as an indication that the gas is not in a state of dynamical equilibrium in those cases, showing also signs of multiple and decoupled kinematic components \citep[see e.g.][]{North21,Oosterloo24}. When the molecular gas in these systems is detected in absorption against a bright and compact background continuum source, evidence have been found that this traces inflowing cold molecular clouds that are moving toward the central SMBH with velocities up to $550$~km~s$^{-1}$, thus likely contributing to the fuelling of the jetted-AGN \citep[see e.g.][]{Tremblay16,Maccagni18,Rose19,Rose23,Rose24}.

In summary, while it is true that the bulk of the molecular gas detected in the majority of local ETGs is regularly-rotating and dynamically cold, various localised levels of perturbations in the gas physics (i.e.\,high excitation conditions), morphology (e.g.\,warps, asymmetries and/or disruptions) and kinematics (i.e.\,inflow/outflow motions) are ubiquitously detected in jetted-AGN hosts. This clearly indicates that - when the central SMBH is caught in an active phase of its life - the gas is not fully relaxed into the host galaxy potential. All this adds also interesting clues on the potential connection between AGN activity and the surrounding (sub-)kpc scale molecular gas discs and, more in general, on the interplay between the AGN and its host galaxy.

\subsection{Gas disc stability: star formation and black hole accretion}
The molecular phase of the ISM is associated to active star formation, so that a rotating molecular gas disc/ring is generally expected to collapse into star-forming clouds. One may thus wonder how star formation proceeds in the gas-rich ETGs discussed above. 
To first order, these appear totally ``dead'' in - e.g.\,- optical colour-magnitude diagrams, because of their prominent red optical colours that barely differ from fully passive systems (see also Section~\ref{sec:intro}). However, by using tracers that are more sensitive to star forming regions (e.g.\,UV emission), some low-level star formation can be picked out also in these systems \citep[e.g.][]{Crocker11,Young14}. The specific star formation rates ($sSFR \equiv SFR/M_{\filledstar}$) typically remain very low in such ETGs, because the gas reservoirs are small and embedded in a massive galaxy \citep[e.g.][]{Davis22}. They also have very low star formation efficiencies ($SFE \equiv SFR/M_{\rm H2}$), for the obvious reason that the rates at which their molecular reservoirs form stars is modest compared to their masses \cite[e.g.][]{Davis14a,Davis22}. All this contributes to ensure that gas-rich ETGs can still be considered overall ``red and dead''. 

The origin of this lowered star formation efficiency in gas-rich ETGs is currently unclear. In general, the fate of cold gas distributions is governed by the balance between mechanisms which tend to gather molecular gas clouds together (such as self-gravity) and dissipative forces (e.g.\,shear and turbulence) which rather tend to pull them apart. As extensively discussed above, energetic feedback from an AGN (either in radiative or kinetic form) can be responsible for changing the physical conditions of the surrounding ISM, by inducing shocks and turbulence, heating, compressing, or even expelling it from the nuclear regions through powerful outflows \citep[e.g.][]{Combes17,Harrison18,Morganti20b}. This can either prevent gas fragmentation and thus successive bursts of star formation (negative feedback) or locally promote it (positive feedback; see e.g.\,\citealp{Harrison24}). Thus, in the current scenario, AGN can play a crucial role in regulating the star formation activity of local gas-rich ETGs \citep[e.g.][]{Choi15}. At the same time, however, it has been shown that other factors may also intervene in these systems and influence the ability of their molecular gas to form stars (also because not all the SMBHs at the centre of local ETGs are currently caught in an active phase of their life). 
 
 ETGs are bulge-dominated systems, having very deep potential wells and thus very centrally-concentrated stellar mass distributions. The circular velocity curves of this type of galaxies therefore rise sharply (as visible also from the observed molecular gas velocity curves; see e.g.\,Figure~\ref{fig:CO_kin} and \ref{fig:jetted_AGN_kin}), leading to strong shear rates. Both observations and simulations have now shown that this, together with the strong gravitational influence exerted in the inner regions by the high-mass SMBHs typically hosted at the centre of ETGs (see Section~\ref{sec:dyn_BH_masses}), have the effect of stabilising the cold gas against gravitational fragmentation, thus impacting the star formation efficiency in these systems \citep[see e.g.][]{Martig13,Davis14b,Boizelle17,Davis22,Williams23}. 

Historically, the local gas state against gravitational fragmentation has been quantitatively inferred from the analysis of the Toomre parameter \citep{Toomre64}:

\begin{eqnarray}\label{eq:Q_param}
Q = \dfrac{\sigma_{\rm gas}\kappa}{\pi G \Sigma_{\rm gas}}
\end{eqnarray}
where $\sigma_{\rm gas}$ and $\Sigma_{\rm gas}$ are the velocity dispersion and surface density, respectively, of the molecular gas, $G$ is the gravitational constant, and $\kappa$ is the epicyclic frequency (i.e.\,the frequency at which a gas parcel oscillates radially along its circular orbit). The latter is calculated as  
\begin{eqnarray}\label{eq:kappa}
\kappa = \sqrt{4\Omega^{2}+R\dfrac{d\Omega^{2}}{dR}}
\end{eqnarray}
where $R$ is the radius of the gas distribution and $\Omega$ is the angular frequency ($\Omega=v_{\rm circ}/R$, with $v_{\rm circ}$ being the de-projected circular velocity). Theoretically, gas discs with Q values above unity are considered stable against gravitational fragmentation, unstable otherwise. In the majority of the local ETGs where this type of analysis has been carried out, the average value of $Q$ is found to be $>1$ \citep[e.g.][]{Davis17,Boizelle17,Voort18,Ruffa19b}. This, in first instance, provides some further support to a scenario in which the deep potentials of massive early-type hosts overall stabilise the gas distributions against gravitational collapse. However, caution is needed in drawing conclusions from the Toomre criterion alone, as $Q>1$ have been measured in some nearby star-forming (disc) galaxies \citep{Romeo17}. Numerical simulations also show that gravitational instabilities cannot be completely excluded in regions where $Q$ is just slightly above unity \citep{Li05}. Furthermore, observational and projection effects (such as those affecting the reliability of observed $\sigma_{\rm gas}$ values; see discussion at the beginning of Section~\ref{sec:gas_kinematics}) may also affect the estimation of $Q$ \citep[see e.g.][]{Ruffa19b}. 
Therefore, a comprehensive analysis of the gas stability state and - more in general - of its star formation abilities requires highly-resolved (cloud-scale) CO observations \citep[see e.g.][]{Williams23}, as well as complementary data (such as spatially-resolved SFR maps) to directly detect star-forming regions.

Finally, we note that a stabilised gas state may also have important implications for the fuelling of the nuclear activity. High mass transfer rates and thus efficient SMBH accretion are indeed expected only if the gaseous circumnuclear matter is gravitationally unstable \citep{Wada92}. The fact that the molecular gas detected on the (sub-)kpc scales of local ETGs is overall found to be stable against gravitational fragmentation may thus also explain the low (or very low) rates of matter accretion onto the central SMBHs of some jetted-AGN hosts \citep[such as LERGs; see e.g.][]{Ruffa19a,Ruffa19b,Ruffa22,Ruffa23}.

\section{Measuring SMBH masses from CO kinematics}\label{sec:dyn_BH_masses}
The last three decades of observations have demonstrated that the mass of central SMBHs ($M_{\rm BH}$) correlates with a number of host galaxy properties, such as the bulge mass/luminosity \citep[e.g.][]{Magorrian98,Lasker14} and the central stellar velocity dispersion (this latter being commonly indicated as the $M_{\rm BH}-\sigma_{\filledstar}$ relation; \citealp[e.g.][]{Ferrarese00,Gebhardt00,Gultekin09,McConnell13,Vandenbosch16}). This has been generally interpreted as implying a sort of self-regulated co-evolution between SMBHs and their host galaxies \citep[e.g.][]{Kormendy13}. In the current scenario, AGN feedback is typically invoked as a crucial element to set up such co-evolution, being responsible for altering the physics and kinematics of the ISM as described in Section~\ref{sec:mol_gas_in_ETG}. The many details of the mechanisms driving and regulating the SMBH-host galaxy interplay, however, are still poorly understood and thus hotly debated \citep[see e.g.][for a recent review]{Donofrio21}.

One of the crucial steps to advance in our understanding of co-evolution is to pin down the SMBH-host galaxy correlations, as well as their true amount of astrophysical scatter and its drivers. However, our ability of making such progresses strongly depends on the accuracy of the $M_{\rm BH}$ measurements. The most reliable estimates are obtained through dynamical studies of matter orbiting within the sphere of influence (SOI) of the SMBH. This is the region inside which the gravitational potential of the SMBH dominates over that of the host galaxy, and is typically defined as $R_{\rm SOI} = \dfrac{GM_{\rm BH}}{\sigma_{\rm \filledstar}^{2}}$, where $G$ is the gravitational constant and $\sigma_{\rm \filledstar}$ is the stellar velocity dispersion within one effective radius \citep[see e.g.][]{Davis14a}. Up until $\sim10$ years ago, dynamical $M_{\rm BH}$ estimates were exclusively obtained by modelling the stellar \citep[e.g.][]{Cappellari02b,Kraj09}, ionised gas \citep[e.g][]{Ferrarese96,Sarzi01,Dallabonta09,Walsh13}, or megamaser kinematics \citep[e.g.][]{Miyoshi95,Greene10,Kuo11} within $R_{\rm SOI}$. 
Studies of the stellar kinematics can provide quite accurate $M_{\rm BH}$ estimates (with typical $1\sigma$ uncertainties $\lesssim10$\%), but they can also be strongly affected by dust extinction and require very high resolution to reliably probe the line-of-sight stellar velocity distribution within the SOI. Dynamical measurements from ionised gas kinematics would be technically simpler, but they are often (if not always) challenged by the presence of non-gravitational forces (e.g.\,shocks) and/or turbulent gas motions superimposed on (quasi-)circular motion. Very high precision is usually achieved when using megamasers as tracers, as these typically probe material very close to the SMBH. However, megamaser emission is usually detected only in certain types of AGN (mostly Seyfert 2, typically in late-type hosts) and only if the megamaser disc is edge-on.

\begin{figure}[]
\begin{adjustwidth}{-\extralength}{0cm}
\centering
\includegraphics[clip=true, trim={1 5 1 5}, scale=0.65]{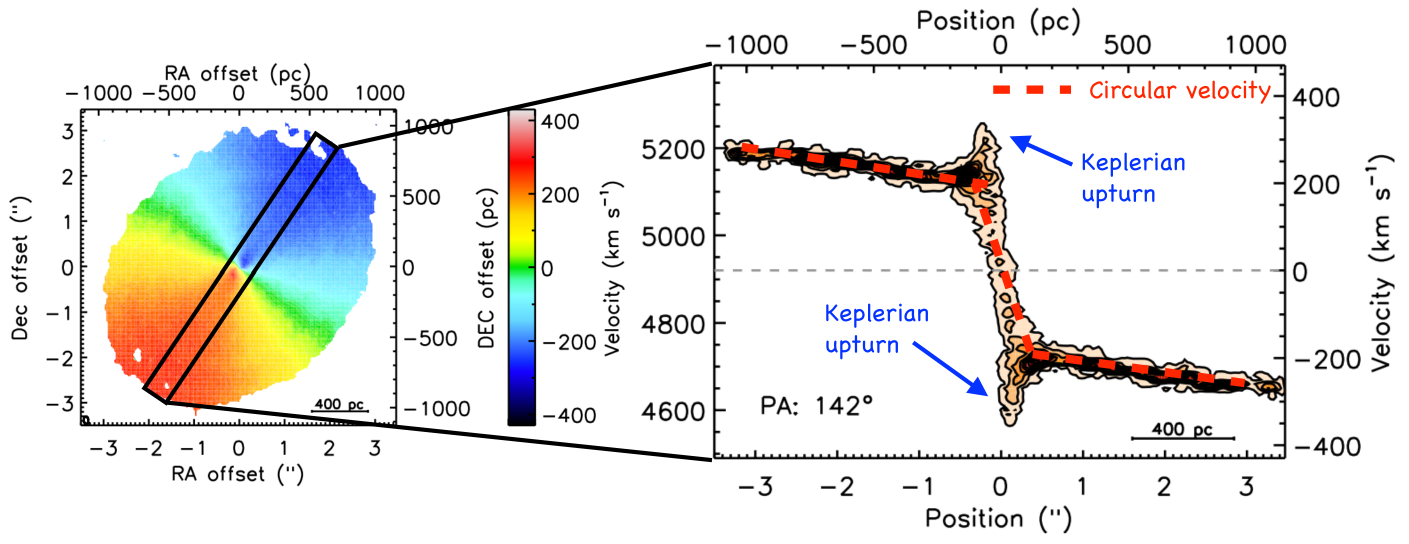}
\end{adjustwidth}
\caption{Cartoon featuring the CO(2-1) moment 1 and PVD maps of the jetted-AGN host NGC\,383. Here the resolution of the ALMA data allows to well resolve the $R_{\rm SOI}$ and thus clearly detect the characteristic Keplerian upturn (indicated by the blue arrows) of matter orbiting around a massive dark object. The maps are reproduced with permission from \citet{North19}.\label{fig:Keplerian_turn}}
\end{figure}

Over the past decade, thanks to the unprecedented resolution and sensitivity provided by ALMA, a new method for SMBH mass measurements has been developed: probing the kinematics of the molecular gas down to the SOI using CO emission lines \citep[e.g.][and references therein]{Davis13a,Barth16,Onishi17,Davis17,North19,Boizelle19,Smith19,Smith21b,Boizelle21,Cohn21,Kabasares22,Ruffa23,Dominiak24a,Dominiak24b,Kabasares24}. The basic assumption of this technique is that the kinematics of the molecular gas within the SOI mainly arises from the gravitational influence of both the luminous stellar component and the SMBH, this latter in particular giving rise to the characteristic Keplerian motions of matter orbiting around it (see Figure~\ref{fig:Keplerian_turn}). Spatially resolving the Keplerian region thus allows to constrain the SMBH mass to high accuracy, once the contribution of visible matter (i.e.\,the gas circular velocity component, red dashed line in Figure~\ref{fig:Keplerian_turn}) is removed from the the observed gas kinematics. A very recent work demonstrates that molecular gas observations of nearby galaxies using the most extended ALMA configurations can penetrate deep within the SOI, thus enabling SMBH mass estimates as precise as the best megamaser ones \citep{Zhang24}.

\begin{figure}[]
\centering
\includegraphics[clip=true, trim={2 4 2 5}, scale=0.5]{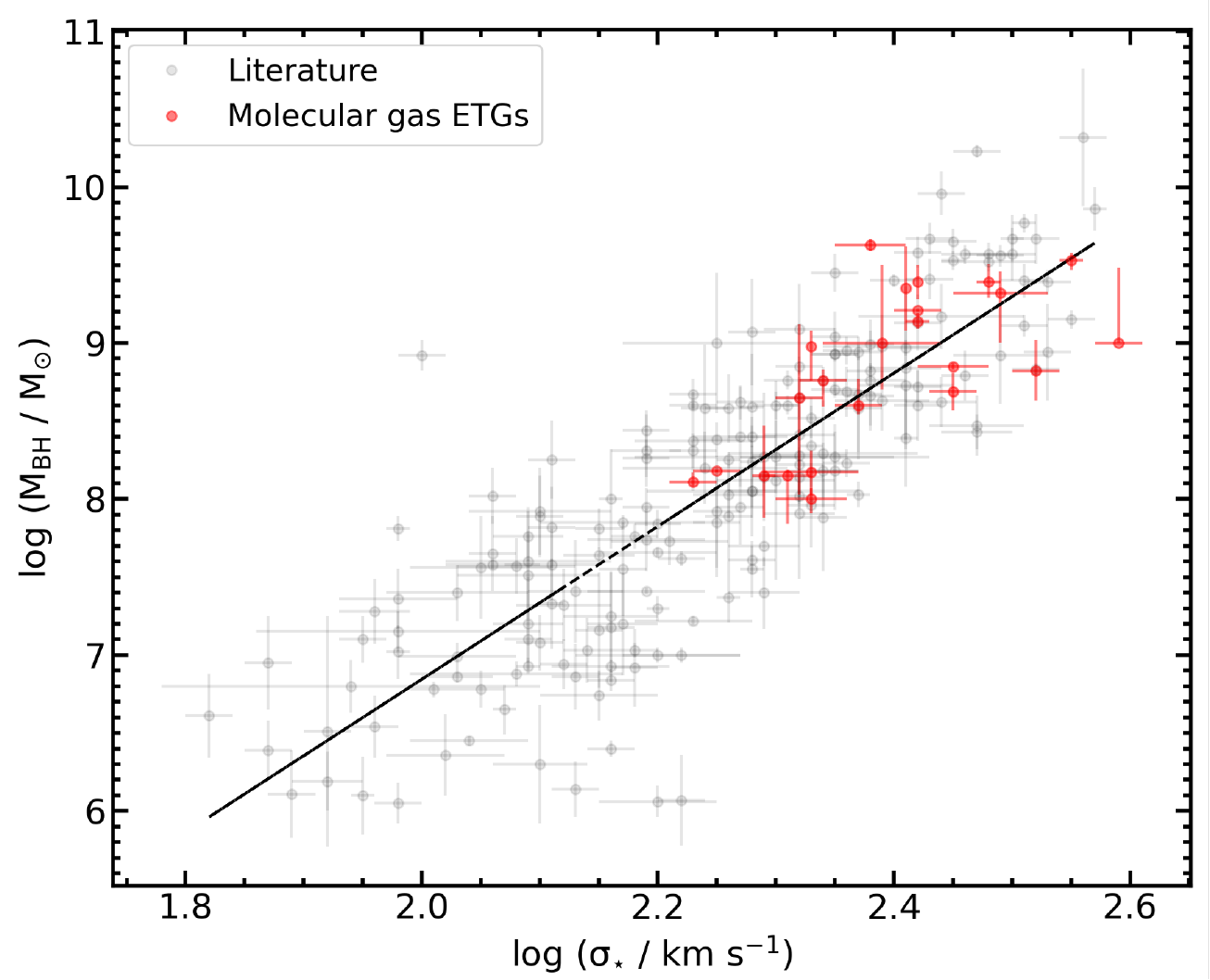}
\caption{$M_{\rm BH}-\sigma_{\rm \filledstar}$ relation from the compilation of \citet[][]{Vandenbosch16} (grey data points and black line). The SMBH mass measurements obtained in nearby ETGs from CO dynamical modelling are overlaid as red dots, and taken from \citealt[][]{Davis13a,Onishi15,Barth16,Onishi17,Davis17,Davis18,Smith19,North19,Boizelle19,Ruffa19b,Boizelle21,Smith21b,Cohn21,Kabasares22,Ruffa23,Dominiak24a,Dominiak24b,Kabasares24}. For CO dynamical SMBH measurements, the error bars correspond to $1\sigma$ uncertainties.\label{fig:M_sigma}}
\end{figure}

This technique has already been successfully applied to different types of galaxies (late- and early-types, active and inactive). However, given the molecular gas properties described in Section~\ref{sec:mol_gas_in_ETG} (i.e.\,bulk of the gas regularly-rotating and dynamically cold in most of the cases), ETGs are ideal targets to use the CO technique to dynamically-measure SMBH masses with high accuracy. The presence of CO morphological asymmetries (e.g.\,warps) and/or nuclear gas deficiencies can be problematic, but it has been already demonstrated that reliable mass estimates can be obtained also in cases like these \citep[e.g.][]{Davis13a,Onishi17,Davis18,Smith19,Ruffa23}. Measuring SMBH masses using high-resolution ALMA CO observations in a diverse sample of galaxies is the primary aim of the mm-Wave Interferometric Survey of Dark Object Masses (WISDOM) project, which has to date provided accurate estimates in ten typical ETGs of all types \citep[][]{Davis13a,Onishi17,Davis17,Davis18,Smith19,North19,Smith21b,Ruffa23,Dominiak24b}. Other research groups carried out SMBH mass estimates for thirteen additional ETGs \citep[][]{Onishi15,Barth16,Boizelle19,Nagai19,Ruffa19b,Boizelle21,Cohn21,Kabasares22,Dominiak24a,Kabasares24}. This is enabling to either obtaining accurate first constraints on the $M_{\rm BH}$ of these objects or robustly cross-checking existing measurements which were previously derived using one of the different techniques mentioned above \citep[see e.g.][]{Boizelle21,Ruffa23}. 

In Figure~\ref{fig:M_sigma} we present the $M_{\rm BH}-\sigma_{\rm \filledstar}$ relation from the compilation of \citet{Vandenbosch16}, with overlaid as red dots the 23 SMBH mass measurements obtained to date applying the molecular gas technique in ETGs. This plot overall shows that, while the majority of ETGs lie well within the scatter of the best-fitting relation (black line in Figure~\ref{fig:M_sigma}), a few massive ones host SMBHs that lie towards the outer edge (or even slightly outside) of such scatter. This has been dubbed as the ``over-massive'' SMBH population \citep[e.g.][]{vandenBosch12}. Galaxies within this class are thought to be local analogues of the higher-redshift quiescent galaxies that also contain over-massive black holes, and could therefore be relics that have evolved little since a redshift $z \approx 2$ \citep[e.g.][]{Vandenbosch16}. Alternatively, an over-massive SMBH may be the result of merger(s), especially when paired with a high molecular gas mass in an ETG hosting an AGN. It is still not clear which of these scenarios best explains the observed behaviour. A recent work, however, suggests that over-massive SMBHs may in fact be fairly uncommon ($\lesssim30\%$; \citealp{Dominiak24a}). Further work focusing on increasing the number of massive ETGs with accurate CO dynamical SMBH mass measurements will enable better constraints on the incidence of this population and - more in general - help us making further progresses on our understanding of the high-mass end of the SMBH-host galaxy correlations. 

\section{Summary and Future Perspectives}
In this work, we have briefly discussed the main lessons learned from the first ten years of ALMA observations of the molecular gas component in local ETGs, restricting our analysis to objects within $z\lesssim0.05$ and making clear distinctions between normal (inactive) ETGs, jetted-AGN hosts and BCGs (typically hosting also a jetted-AGN). The main takeaway points from our short review can be summarised as follows:
\begin{itemize}
    \item {\bf Molecular gas content and distribution.} Molecular gas reservoirs with masses ${\gtrsim10^{7}}$~M$_{\odot}$ are found in $\approx$25\% of local ETGs. In the majority of normal ETGs and jetted-AGN hosts (including some BCGs), these significant amounts of gas have been observed to be distributed in smooth, thin disc-like structures, extending on scales from a few hundred parsecs to a few kpc around the nuclear galaxy regions. In $\approx20$\% of the cases, the gas presents circumnuclear holes, with sizes ranging from few tens to few hundreds of parsecs. The origin of these nuclear gas deficiencies is still debated (they may be due to either radiative dissociation, high gas excitation conditions, dynamical or mechanical effects). In the majority of BCGs the same large amounts of molecular gas are typically detected in the form of kpc-scale filamentary or blob-like structures, which are consistent with expectations from cooling flows.
    \item {\bf Molecular gas kinematics.} In normal ETGs, the molecular gas is typically found to be regularly rotating and dynamically cold (with average $\sigma_{\rm gas} \sim 10-20$~km~s$^{-1}$).  Departure from symmetry (such as inclination or velocity warps) and/or secondary kinematic components (such as inflow/outflow motions) are rare. This is generally interpreted as an indication of gas that is both dynamically and morphologically settled within the potential well of the host galaxy. In jetted-AGN hosts, while the bulk of the molecular gas is still consistent with being regularly-rotating and dynamically cold, various levels of local perturbations in the gas physics (i.e.\,high excitation conditions), morphology (e.g.\,warps, asymmetries and/or disruptions) and kinematics (i.e.\,inflow/outflow motions) are ubiquitously observed. In general, this indicates that - when the central SMBH is caught in an active phase of its life - the gas is not fully relaxed into the host galaxy potential. In a few cases, by carrying out a detailed multi-wavelength analysis (including ALMA observations of multiple molecular gas tracers), it has been clearly shown that such alterations have to be ascribed to a jet-ISM interaction on (sub-)kpc scales. Clear evidence of cold gas inflows onto galactic centres have been also found in the BCGs with gas distributed in filamentary or blob-like structures, where the gas overall tends to show complex kinematics (with little or no evidence of regular rotation). 
    \item {\bf Gas disc stability.} Even when hosting large amounts of cold molecular gas onto their central regions, local ETGs remain overall “red and dead”. While AGN feedback (either in radiative or kinetic form) can clearly play a role in maintaining them in this state, both simulations and high-resolution (i.e.\,few tens of pc) ALMA observations have now shown that it is likely the strong shear rates induced by their deep potentials the dominant mechanism which stabilises the gas against gravitational collapse (thus hampering the formation of new stars). The fact that the molecular gas detected on the (sub-)kpc scales of local ETGs is overall found to be stable against gravitational fragmentation may also explain the low (or very low) rates of matter accretion onto the central SMBHs of some jetted-AGN hosts (such as LERGs).
    \item {\bf CO dynamical SMBH masses.} Before the advent of ALMA, dynamical estimates of the SMBH mass relied exclusively on studies of the stellar, ionised gas and megamaser kinematics within the SMBH gravitational sphere of influence ($R_{\rm SOI}$). Over the past decade, thanks to the ability of the most extended ALMA configurations to penetrate deep within the SOI, a new method for SMBH mass measurements has been developed and successfully applied to a varied range of galaxy types: probing the Keplerian motion of the molecular gas within the SOI using CO emission lines. Thanks to their molecular gas properties, ETGs are ideal targets for these type of studies. This is allowing us to put solid constraints on the high-mass ends of SMBH-host galaxy correlations (such as the $M_{\rm BH}-\sigma_{\rm \filledstar}$), and will ultimately help us making further progresses in our understanding of the SMBH-host galaxy interplay (so-called co-evolution).
\end{itemize}

In short, it is clear that the first ten years of ALMA observations enabled transformative studies of the cold molecular phase of the ISM in nearby ETGs, allowing us to map its distribution and kinematics with unprecedented detail and thus to reveal some crucial aspects of the complex physics governing its survival and fate. This is in turn adding essential pieces of knowledge to our understanding of local galaxy populations and of the dark monsters hosted at their centres, and will ultimately help us to shed light on some long-standing debates about their observed properties. 

Thanks to the latest- and next-generation radio and optical telescopes, whose exquisite capabilities complement those of ALMA in the (sub-)mm, we are already in a new exciting era for these type of studies. A comprehensive view of the cold ISM in local ETGs indeed requires also gathering detailed information on the atomic gas and the dust components. The former provides the material from which the molecular medium can form and is typically traced by the atomic hydrogen (HI) hyper-fine transition at wavelengths of 21~cm ($1.4$~GHz). This can be found in both the circumnuclear and circumgalactic regions of ETGs in the form of diffuse structures which can have very low column densities (i.e.\,N$_{\rm HI}\lesssim10^{19}$~cm$^{-2}$) and therefore require very high sensitivities to be detected. The south-african SKA pathfinder (MeerKAT) is currently the best-suited instrument for these kind of studies, and is already allowing major steps forward in the field \citep[see e.g.][]{Maccagni21}. Thanks to its 1.35x and 2x higher sensitivity and resolution, respectively, the planned MeerKAT extension (known as MeerKAT$+$) will soon allow to achieve unprecedented performances in both cases, laying the ground for the full Square Kilometre Array (SKA) revolution. 

Cold dust grains and - in particular - the polycyclic aromatic hydrocarbons (PAHs) play a crucial role in a variety of processes within the ISM. For instance, they provide the surface for the formation of the molecular hydrogen (H$_2$) and are responsible for the photo-electric heating of neutral gas. PAHs also absorb up to 20\% of the UV light from young stars, re-emitting it in the near- and mid-infrared. This makes them promising, alternative tracers of star formation and the total molecular gas mass. Furthermore, when spatially-resolved information on the cold dust are compared with those of the molecular gas, it is possible to set crucial constraints on the origin of the cold ISM \citep[see e.g.][]{Ruffa19b,Ruffa22}. This - together with the information obtained from HI studies - can provide a full, comprehensive picture of the cold ISM in massive quiescent galaxies. The James Webb Space Telescope (JWST) has already made strides in advancing our understanding of the dust component in nearby star-forming galaxies, and is expected to allow the same advances also for ETGs (where the cold dust properties are currently poorly constrained). 

Finally, it is also worth mentioning that - in about ten years from now - the next-generation Very Large Array (ngVLA) is expected to start its full scientific operations. This will enable studies in the local Universe that will be complementary to those of both ALMA and SKA in terms of frequency coverage, but with much higher sensitivity and resolution ($\sim10$x higher sensitivity at 3~mm and baselines $\sim60$x longer than ALMA). At the same time, it will allow us to reveal the details of the molecular gas component in the high-redshift (i.e.\,${z\gtrsim1}$) counterpart of local ETGs and their progenitors DSFGs, therefore making possible a full comprehension of the evolutionary track of massive spheroids. 

All of the above will allow us to build up an increasingly complete observational picture of red sequence galaxies. Once meaningfully coupled with high-resolution simulations and appropriate theoretical models, this will clearly enable major steps forward in our understanding of this galaxy population. 

\vspace{6pt} 




\authorcontributions{IR and TAD contributed jointly to the Literature research underlying this work and to the writing of the manuscript.}

\funding{The research underlying this article was funded by grant \#ST/S00033X/1 through the UK Science and Technology Facilities Council (STFC).}



\dataavailability{All of the data discussed in this review are available to download at the ALMA Science Archive (\url{https://almascience.nrao.edu/asax/}).} 

\acknowledgments{IR would like to thank Matteo Bonato and Marcella Massardi (Galaxies's Guest Editors for this Special Issue) for inviting this review, which is mostly based on the contribution IR presented at the ``Fifth Workshop on Millimetre Astronomy'' (Bologna, IT, May 2023). This paper makes use of ALMA data. ALMA is a partnership of ESO (representing its member states), NSF (USA) and NINS (Japan), together with NRC (Canada), NSC and ASIAA (Taiwan), and KASI (Republic of Korea), in cooperation with the Republic of Chile. The Joint ALMA Observatory is operated by ESO, AUI/NRAO and NAOJ. This paper has also made use of the NASA/IPAC Extragalactic Database (NED) which is operated by the Jet Propulsion Laboratory, California Institute of Technology under contract with NASA. We acknowledge also the usage of the HyperLeda database (\url{http://leda.univ-lyon1.fr}).}

\conflictsofinterest{The authors declare no conflict of interest.} 




\begin{adjustwidth}{-\extralength}{0cm}

\reftitle{References}


\bibliography{mybibliography}

\PublishersNote{}
\end{adjustwidth}
\end{document}